\documentstyle[aps,preprint]{revtex}
\begin{document}
\draft
\title{The quantization of the chiral Schwinger model based on \\
the BFT-BFV formalism II}

\author{Mu-In Park$^1$\footnote{Electronic address: 
 mipark@physics.sogang.ac.kr},
    Young-Jai Park$^1$\footnote{Electronic address: yjpark@ccs.sogang.ac.kr},
    and Sean J. Yoon$^2$\footnote{Electronic address: yoonsj@mail.lgcit.com}}
\address{$^1$Department of Physics and Basic Science Research Institute \\
     Sogang University, C.P.O. Box 1142, Seoul 100-611, Korea \\
     and \\
     $^2$LG Corporate Institute of Technology, Seoul 137-140, Korea}

\maketitle
\begin{abstract}
We apply an improved version of Batalin-Fradkin-Tyutin (BFT) Hamiltonian method 
to the $a=1$ chiral Schwinger Model, which is much more nontrivial
than the $a>1$ one.
Furthermore, through the path integral quantization, 
we newly resolve the problem of the non-trivial $\delta$ function
as well as  that of the unwanted Fourier parameter $\xi$ in the measure.
As a result, we explicitly obtain
the fully gauge invariant partition function,
which includes a new type of Wess-Zumino (WZ) term 
irrelevant to the gauge symmetry as well as usual WZ action.
\end{abstract}


\newpage

\section{Introduction}

After Dirac's pioneering work [1] on the constraint system,
Batalin, Fradkin, and Vilkovsky (BFV) [2] had 
proposed a new kind of
quantization method for constraint systems, 
which is particularly powerful for deriving
a covariantly gauge-fixed action in configuration space.
This BFV formalism has been applied to several interesting models [3]
including the bosonized Chiral Schwinger Model (CSM) [4].
After BFV's work,
Batalin-Fradkin-Tyutin (BFT) [5,6] had generalized this method
for systems with only second class or both
classes of constraints.
Recently, several authors have systematically applied the BFT
Hamiltonian method to several Abelian second class systems [7--9]
finding the new type of an Abelian Wess-Zumino (WZ) action.
After their works, we have also quantized other several interesting
models [10,11] by using this BFT formalism. However, these works [7--11] are
mainly based on the systematic, but somewhat cumbersome construction
of the first class Hamiltonian as a solution of commuting 
with the first class constraints.

Very recently,
we have improved the usual BFT method by obtaining the desired first class
Hamiltonian from the canonical Hamiltonian just by replacing the original fields
with BFT physical fields for the several models [12,13] including 
the $a>1$ case of the CSM having two second class constraints[14]. 
Here, these BFT fields are obtained from the requirement of commuting with the first class constraints.
On the other hand, up to now previous works of the highly nontrivial $a=1$ CSM 
having four second class constraints, which is in contrast with the simple $a>1$ CSM
having only two second class ones, are still problematic in this direction [7, 11]. 
The final quantum theory in those works seems
not to have the prevailing gauge symmetry contrast to the $a>1$ case although we make the second class constraint system 
into the first class one following BFT method.

In this paper, we shall newly resolve this unsatisfactory situation of the $a=1$ CSM based on our improved BFT 
method[12-14] and the non-trivial application
of the recently proposed technique of covariant path integral evaluation [3]. 
In fact, we find that the final quantum action in a much simpler form and 
the final configuration space partition function has the prevailing gauge symmetry contrast to the previous works [7, 11] in conformity with the
effectively first class constraint structure of the model.

The organization of the paper is as follows:
In section II, we convert all second class constraints of the bosonized
CSM with $a=1$ into the effectively first class constraints
according to the usual BFT formalism.
In section III, according to our recently improved BFT method, we then
directly obtain the desired first class Hamiltonian
from the canonical Hamiltonian by simply replacing the original fields
with the corresponding BFT physical variables defined
in the extended phase space
and recover the Dirac bracket (DB) in the original phase space.
In section IV, we obtain the first class quantum Lagrangian,
which includes new type of WZ (NWZ) action which can not be obtained in the
usual Faddeev-Popov like path integral framework [16] as well as the well
known WZ terms, and the final theory has the prevailing gauge symmetry
contrast to the previous works [7, 11].
This becomes possible by resolving the problem of the non-trivial
$\delta$-function in the measure part, which exists 
in the previous works [7,11].
Furthermore, according to our procedure, the Fourier parameter $\xi$
introduced when exponentiating the delta function
$\delta(\tilde{\Omega}_2)$ is aiso disappeared in the final result.
Section V is devoted to conclusion.

\section{Construction of First Class Constraints }

As is well-known, the (fermionic) CSM model is equivalent to the following bosonized action 
\begin{equation}
    S_{CSM} ~=~ \int d^2x~\left[
                          -\frac{1}{4}F_{\mu \nu}F^{\mu \nu}
                          +\frac{1}{2}\partial_{\mu}\phi\partial^{\mu}\phi
                          +eA_{\nu}(\eta^{\mu \nu}
                          -\epsilon^{\mu \nu})\partial_{\mu}\phi
              +\frac{1}{2}a e^{2}A_{\mu}A^{\mu}~\right] ,
\end{equation}
where $\eta^{\mu\nu}=\mbox{diag(1,-1)}$, $\epsilon^{01}=1$, and $a$ 
is a regularization ambiguity, which is not uniquely determined by the 
different procedures for calculating the fermionic determinant [4]. 
Depending on the value of $a$, three different mass spectrums are expected : 
massless harmonic excitations $(a=1)$, massive vector meson with massless 
harmonic excitations $(a>1)$, tachyonic excitations $(a<1)$. 
In addition, these differences are reflected also in the 
constraint structures: four second class constraints $(a=1)$ or two second 
class ones $(a>1)$, where we neglect the non-unitary case of $a<1$. 
Because of the doubling of the number of constraints, 
$a=1$ CSM was highly non-trivial, and thus 
previous works in the BFT Hamiltonian method were still problematic . 
Since resolving the unsatisfactory situation of $a=1$ CSM is our main issue,
we will only consider $a=1$ CSM from now on. 

In the case of $a=1$, the action (1) has the gauge anomaly 
$\delta S_{CSM} =-\int d^2 x e^2 \Lambda^* F$ under 
the gauge transformation $\delta A_{\mu} =\partial_{\mu} \Lambda,~ \delta \phi 
=-e \Lambda$ where ${}^{*}F=\frac{1}{2} \epsilon_{\alpha \beta} F^{\alpha \beta}$, 
dual to $F^{\mu \nu}$.
The canonical momenta are given by
\begin{eqnarray}
    \pi^\mu~&=&~(0, ~\dot{A}_1~-~\partial_{1}A_0),\nonumber \\
    \pi_\phi~&=&~\dot{\phi}~+~e(A_0~-A_1),
\end{eqnarray}
where the overdot denotes the time derivative.
Following Dirac's standard procedure [1,17],
one finds one primary constraint
\begin{equation}
    \Omega_1 \equiv \pi^0 \approx 0,
\end{equation}
and three secondary constraints
\begin{eqnarray}
    \Omega_2 &\equiv& \partial_1 \pi^1 + e\pi_\phi + e\partial_1 \phi
              + e^2 A_1 , \nonumber \\
    \Omega_3 &\equiv& e^2 \pi^1 , \nonumber \\
    \omega_4 &\equiv& e^4(A_0-A_1) + e^2 \partial_1 \pi^1 -e^2 \Omega_2.
\end{eqnarray}
These constraints are obtained by conserving the
constraints with respect to the total Hamiltonian
\begin{equation}
    H_T = H_c + \int dx~u\Omega_1 ,
\end{equation}
where $u$ denotes a Lagrange multiplier. 
Here $H_c$ is the canonical Hamiltonian
\begin{eqnarray}
    H_c &=& \int\!dx~\left[~
            \frac{1}{2}(\pi^{1})^2 + \frac{1}{2}(\pi_{\phi})^2
             + \frac{1}{2}(\partial_{1}\phi)^2 + e(\pi_{\phi}
             + \partial_{1}\phi) A_1 \right. \nonumber \\
        &&~~~~~~ \left. +e^{2}(A_{1})^2 -A_0 \Omega_2~~\right].
\end{eqnarray}
By fixing the Lagrange 
multiplier $u$ as follows 
\begin{equation}
        u =\partial_1 A_0 -\frac{1}{e^4}\partial_1 \omega_4 +\frac{2}{e^2} \Omega_3     ,
\end{equation}
no further constraints are generated via this procedure.
We find that all the constraints are fully second class constraints.
In order to carry out simple algebraic manipulations (i.e. avoiding
the complication which comes from the derivative term 
$\{\omega_4 (x),\omega_4 (y)\} = 2 e^6 \partial^x_1 \delta (x-y) $),
we redefine $\omega_4$ by using $\Omega_1$ as
follows
\begin{eqnarray}
        \Omega_4 &\equiv&
                \omega_4 + e^2 \partial_1 \Omega_1 \nonumber\\
         &=& e^4 (A_0 -A_1) +e^2 \partial_1 \pi^1 -e^2 \Omega_2 +e^2 \partial_1 \Omega_1 .
\end{eqnarray}
Note that the redefined constraint is still
second class one in contrast to the Chern-Simons (CS) theories [8,10].
The full set of the constraints 
$\Omega_i (i=1,\cdot\cdot\cdot,4)$ now satisfies the simplified form of the second class constraints algebra 
\begin{eqnarray}
    \Delta_{ij}(x,y)
                &\equiv&  \{ \Omega_{i}(x), \Omega_{j}(y) \} 
                                            \nonumber\\
                &=& e^4  \left( \begin{array}{cccc}
                                          0   &  0   &  0    &  -1     \\
                                          0   &  0   &  1    &  0          \\
                                          0   & -1   &  0    & 2e^2     \\
                                          1 &  0     & -2e^2 &  0
                                 \end{array} \right)
  \delta(x -y), 
\end{eqnarray}
and furthermore which are closed under the time translation as follows
\begin{eqnarray}
&&\dot{\Omega}_1 =\{ \Omega_1, H_T \}=\Omega_2, \nonumber \\
&&\dot{\Omega}_2 =\{ \Omega_2, H_T \}=\Omega_3+2 e^2 \Omega_1, \nonumber \\
&&\dot{\Omega}_3 =\{ \Omega_3, H_T \}=\Omega_4+e^2 \partial_1 \Omega_1, \nonumber \\
&&\dot{\Omega}_4 =\{ \Omega_4, H_T \}=(2 e^2 \partial_1 \partial_1 -4 e^4)\Omega_1+e^2 \partial_1 \Omega_2
\end{eqnarray}
with the multiplier of (7).

Now, let us introduce new auxiliary fields $\Phi^{i}$
in order to convert the second class constraints $\Omega_{i}$ into
first class ones in the extended phase space. 
Following the usual BFT method [5,6], 
we require these fields to satisfy
\begin{eqnarray}
   \{{\cal F}, \Phi^{i} \} = 0,~~ \{ \Phi^i(x), \Phi^j(y) \} = \omega^{ij}(x,y),
\end{eqnarray}
where $\omega^{ij}$ is a constant and antisymmetric matrix and 
${\cal F} \equiv (A^{\mu},\pi_{\mu},\phi, \pi_{\phi} )$ are
the variables of the original phase space.
Then, the strongly involutive modified constraints
$\widetilde{\Omega}_{i}$, which satisfy
\begin{eqnarray}
\{\widetilde{\Omega}_{i}, \widetilde{\Omega}_{j} \}=0;
~~~\widetilde{\Omega}_i \mid_{\Phi = 0} = \Omega_i,
\end{eqnarray}
are formally given by
\begin{equation}
  \widetilde{\Omega}_i( {\cal F}; \Phi)
         =  \Omega_i + \sum_{n=1}^{\infty} \widetilde{\Omega}_i^{(n)},
                       ~~~~~~\widetilde{\Omega}_i^{(n)} \sim (\Phi)^n.
\end{equation}
Furthermore, we could generally take 
the first order correction terms in the infinite series (13)
as follows [5,6]
\begin{equation}
  \widetilde{\Omega}_i^{(1)}(x) = \int dy X_{ij}(x,y)\Phi^j(y).
\end{equation}
Then, the first class  
constraint algebra of $\widetilde{\Omega}_i$ imposes
the following condition:
\begin{equation}
   \triangle_{ij}(x,y) +
   \int dw~ dz~
        X_{ik}(x,w) \omega^{kl}(w,z) X_{jl}(y,z)
         = 0.
\end{equation}
According to the usual BFT formalism,
we can take without any loss of generality the simple solutions as
\begin{equation}
        \omega^{ij} (x,y) ~=~
                \left( \begin{array}{cccc}
                        0       &  0  &  0  &  1    \\
                        0       &  0  &  1  &  0    \\
                        0       & -1  &  0  &  0    \\
                   -1   &  0  &  0  &  0
           \end{array}
   \right)
   ,
\end{equation}
and
\begin{equation}
X_{ij} (x,y) ~=~
                   e^2 \left( \begin{array}{cccc}
                        -1      &    0     &    0   &    0    \\
                        0       &  -1      &    0   &    0    \\
                        e^2     &    0     &    1   &    0    \\
                        0       &   e^2    &    0   &   -1
                \end{array} \right)
        \delta(x -y)~\equiv X^0_{ij} \delta(x -y) .
\end{equation}
With this choice,
the modified constraints up to first order,
\begin{eqnarray}
\tilde{\Omega}_i =\Omega_i + \tilde{\Omega}^{(1)}_i = \Omega_i + X^0_{ij} \Phi^j
\end{eqnarray}
form a strongly first class  constraint algebra
\begin{equation}
  \{\Omega_{i}+ \widetilde{\Omega}^{(1)}_{i},
            \Omega_{j}+ \widetilde{\Omega}^{(1)}_{j} \} = 0.
\end{equation}
The higher order correction terms 
$\widetilde{\Omega}^{(n)}_{i} ~(n \geq 2)$ in Eq. (13)
automatically vanish as a consequence of the proper choice (16) and (17). 
Therefore, we have all the first class constraints
in the extended phase space
with only $\Omega_i^{(1)}$ contributing in the series (13).

\section{BFT Physical Variables and First Class Hamiltonian}

In this section, we will show that one can easily
find the first class Hamiltonian for the extened phase space only by
replacing the original fields 
${\cal F} = (A^{\mu},{\pi}_{\mu},{\phi}, {\pi}_{\phi} )$
in the canonical Hamiltonian 
with the BFT physical variables
$\widetilde{\cal F} \equiv (\widetilde{A}^{\mu},\widetilde{\pi}_{\mu},
\widetilde{\phi}, \widetilde{\pi}_{\phi} )$
defined in the extended phase space. 
In fact, these BFT physical variables
$\widetilde{\cal F}$ are obtained as a power series
in the auxiliary fields $\Phi$ by requiring them to be strongly
involutive with the first class constraints
\begin{eqnarray}
\{ \widetilde{\Omega}_{i}, \widetilde{\cal F} \} =0.
\end{eqnarray}
The first order correction terms in Eq. (20) are given by
\begin{equation}
\widetilde{\cal F}^{(1)}(x) =  - \int dy \int dz \int dw \Phi^{j}(y) \omega_{jk} (y,z) X^{kl}(z,w)
 \{ \Omega_{l}(w), {\cal F}(x) \}_{({\cal F})}.
\end{equation}
Here, $\omega_{jk}$ and $X_{kl}$ denote the inverse of 
$\omega^{jk}$ and $X^{kl}$.
Furthermore, as a consequence of the proper choice (16) and (17),
the higher order correction terms 
$\widetilde{\cal F}^{(n)} ~ (n \geq 2)$ vanish. 
Hence, the desired BFT physical variables are given by
\begin{eqnarray}
\widetilde{A}^{\mu} &=& A^{\mu} + \widetilde{A}^{\mu (1)} 
                     = (A^{0}  -\partial_{1} \Phi^{1}  - \Phi^{2} -
\frac{1}{e^2}\Phi^{4} ,
      A^{1} + \partial_{1} \Phi^{1}  + \Phi^{2} - 
         \frac{1}{e^2}\partial_{1} \Phi^{3} )  \nonumber \\
\widetilde{\pi}^{\mu} &=& \pi^{\mu} + \widetilde{\pi}^{\mu (1)} 
                      = (\pi^{0}-e^2 \Phi^{1}, 
                         \pi^{1}+e^2 \Phi^{1}+ \Phi^{3}    ) \nonumber \\
\widetilde{\phi} &=& \phi + \widetilde{\phi}^{(1)}  
                  = \phi - \frac{1}{e} \Phi^{3} \nonumber \\
\widetilde{\pi}_\phi &=& \pi_{\phi} + \widetilde{\pi}_\phi^{(1)} 
                    = \pi_{\phi} - \frac{1}{e}\partial_{1} \Phi^{3}.
\end{eqnarray}

On the other hand, the corresponding first class quantities
$\widetilde{K}({\cal F}; \Phi)$ for the arbitrary function $K ({\cal F})$
can be constructed along similar lines 
as in the case of the first class variables,
by representing it as a power series in the auxiliary fields 
and requiring $\{ \widetilde{\Omega}_{i}, \widetilde{K} \} =0$
subject to the condition $\widetilde{K}|_{\Phi =0} =K$.
However, instead of seeking such functions by the usual BFT method,
we consider a new approach using the novel property [12-14] by noting
that any functional of first class BFT fields $\widetilde{\cal F}$ 
corresponding to ${\cal F}$ will also be first class as follows
\begin{eqnarray}
&&\widetilde{K}({\cal F}; \Phi) =
K(\widetilde{\cal F}), \\
&&\{ K(\widetilde{\cal F}),\widetilde{\Omega}_{i} \}=0.
\end{eqnarray}

Now, using this elegant property,
we directly find the first class Hamiltonian $\widetilde{H}_c$
from the canonical one $H_{c}$
only by replacing the original fields with the BFT physical variables as follows
\begin{eqnarray}
   \widetilde{H}_c({\cal F}; \Phi)&=& H_{c}(\widetilde{\cal F}) \nonumber \\
     &=&H_c ({\cal F}) + \int d x \left[ \Phi^1 (-\frac{1}{2} e^4 \Phi^1 
     -e^2 \Phi^3 - \partial_1 \Phi^4 +e^2 \partial_1 A^0 ) 
     + \Phi^2 \Phi^4 -\frac{1}{2} \Phi^3 \Phi^3 \right. \nonumber \\
     ~~&& +\partial_1 (\partial_1 \Phi^1 +\Phi^2 ) \tilde{\Omega}_1 + (\partial_1 \Phi^1 +\Phi^2 +\frac{1}{e^2} \Phi^4 )\tilde{\Omega}_2 +(\Phi^1 + \frac{1}{e^2} \Phi^3 ) \tilde{\Omega}_3 \nonumber \\
     ~~&& \left. +\frac{1}{e^2} (\partial_1 \Phi^1 + \Phi^2 ) \tilde{\Omega}_4  \right]
\end{eqnarray}
and by construction, is strongly involutive 
with the first class constraints (18),
\begin{eqnarray}
\{ \widetilde{\Omega}_{i}, \widetilde{H}_c \} =0.
\end{eqnarray}
Note that the modified constraints have already
the property (23), i.e.,$\widetilde{\Omega}_{i}({\cal F} ; \Phi)=
\Omega_{i}(\widetilde{\cal F})$. 
In this way, all the second class constraints 
$\Omega_{i}({\cal F})$ 
are possible to convert into
the first class ones $\widetilde{\Omega}_{i}({\cal F}; \Phi )$ 
satisfying the boundary conditions
$\widetilde{\Omega}_{i} |_{\Phi=0}=\Omega_{i}$.

Since in the Hamiltonian formalism the first class constraint system
indicates the presence of a local symmetry,
this completes the  operatorial conversion of the original
second class system with Hamiltonian $H_c$ and constraints $\Omega_i$
into the first class ones in the extended phase space
with Hamiltonian $\widetilde H$ and constraints
$\widetilde{\Omega}_i$ by using the property (25).

\section{Path Integral Quantization and Quantum Lagrangian}

In this section, we consider the partition function of the model
in order to extract out the Lagrangian corresponding to $\widetilde{H}$.
To do this, let us identify the new variables
$\Phi^i$ as a canonically conjugate pairs 
\begin{equation}
    \Phi^i \equiv ( \frac{1}{e} \theta,
                  - \rho,
                  - \pi_\rho,
                   e\pi_\theta )
\end{equation}
satisfying Eqs. (11) and (16).
Then, our starting partition function is given by the 
Faddeev-Popov (FP)
formula [16] as follows
\begin{equation}
    Z =  \int  {\cal D} A_\mu  {\cal D} \pi^\mu
               {\cal D} \phi {\cal D} \pi_\phi
               {\cal D} \theta {\cal D} \pi_\theta
                           {\cal D} \rho   {\cal D} \pi_\rho
                 \prod_{i,j = 1}^{4}
                 \delta(\tilde{\Omega}_i )\delta(\Gamma_j )
                 \det \mid \{\tilde{\Omega}_i,\Gamma_j \} \mid
                 e^{iS},
\end{equation}
where
\begin{equation}
        S  =  \int d^2x \left(
                \pi^\mu {\dot A}_\mu
                + \pi_\phi {\dot \phi}
                + \pi_\theta {\dot \theta}
                + \pi_\rho {\dot \rho} - \tilde{\cal H}_c
                   \right)
\end{equation}
with the Hamiltonian density $\widetilde {\cal H}_c$ corresponding to the 
Hamiltonian
${\widetilde H}_c$ of Eq. (25), which is now explicitly expressed in terms
of $(\theta, \pi_\theta, \rho, \pi_\rho)$ instead of $\Phi^i$.
The gauge fixing conditions $\Gamma_i$ should be chosen
so that the determinant occurring in
the functional measure is non-vanishing.
Moreover, $\Gamma_i$ may be assumed to be independent of the momenta
so that these are considered as FP type gauge conditions
[7-14].

Before performing the momentum integrations to obtain the desired
partition function in the configuration space,
it seems appropriate to comment on the strongly involutive Hamiltonian (25).
If we use the above Hamiltonian, we cannot naturally generate
the first class constraints $\tilde{\Omega}_i (i=2,3,4)$ from
the time evolution of the primary constraint $\tilde{\Omega}_1$.
In order to avoid this situation,
we also use another equivalent first class Hamiltonian
without any loss of generality,
which only differs from the involutive Hamiltonian (25)
by the terms proportional to the first class constraint
$\tilde{\Omega}_i$ as follows
\begin{eqnarray}
        \tilde{H}^{*} = \tilde{H}_c + (u-\frac{1}{e^2}\partial_1 \Omega_2)\tilde{\Omega}_1
                  +(\rho - \frac{\pi_\theta}{e} -\frac{1}{e^2} \partial_1 \Omega_1) \tilde{\Omega}_2 
+ (\frac{\theta}{e}+ \frac{\pi_\rho}{e^2} )\tilde{\Omega}_3
                    + \frac{\rho}{e^2} \tilde{\Omega}_4  .
\end{eqnarray}
Then, this Hamiltonian $\tilde {H}^{*}$ automatically generates
the first class constraints
such that $\{ \tilde{\Omega}_1, \tilde{H}^{*} \} =\tilde{\Omega}_{2}, 
\{ \tilde{\Omega}_2, \tilde{H}^{*} \} =\tilde{\Omega}_{3}+2 e^2 \tilde{\Omega}_1,
\{ \tilde{\Omega}_3, \tilde{H}^{*} \} =\tilde{\Omega}_{4}+ e^2 \partial_1 \tilde{\Omega}_1,
\{ \tilde{\Omega}_4, \tilde{H}^{*} \} =(2 e^2 \partial_1 \partial_1 -4 e^4) \tilde{\Omega}_1 +e^2 \partial_1 \tilde{\Omega}_{2}$
in exactly the same way as $H_T$ does $\Omega_i$ of (10).
There is another reason in using the Hamiltonian $\tilde {H}^{*}$ instead of 
$\tilde {H}_{c}$. 
As we have mentioned, all the theories for the extended phase space can be obtained
just by replacinging the original variables ${\cal F}$ 
with the BFT physical variables $\tilde{\cal F}$ in the theories
for the original phase space.
In this point of view, the equations of motion for $\tilde{\cal F}$ 
obtained by a) using the equation 
$\dot{\tilde{\cal F}}=\{\tilde{\cal F},\tilde{H}\}$ 
and by b) replacing ${\cal F}$ with $\tilde{\cal F}$ in the equations of motion
for ${\cal F}$ should be the same.
Interestingly enough, not $\tilde {H}_{c}$ but  $\tilde {H}^{*}$ 
realizes this requirement, in other words the latter
gives form-invariant equations of motion.   
Consequently the first-class Hamiltonian $\tilde {H}^{*}$ for the extended phase space
is made unique by requiring that it reproduces the Heisenberg equations of motion. 
However, we note that $\widetilde{H}_c$ and $\widetilde{H}^*$ act in the same way
on physical states because such states are annihilated by the first 
class constraints which are the only difference between 
$\tilde{H}_c$ and $\tilde{H}^*$. 

Now, we consider the following effective phase space
partition function
\begin{eqnarray}
    Z &=&  \int  {\cal D} A_\mu  {\cal D} \pi^\mu
               {\cal D} \phi {\cal D} \pi_\phi
               {\cal D} \theta {\cal D} \pi_\theta
                           {\cal D} \rho   {\cal D} \pi_\rho
                 \prod_{i,j = 1}^{4}
                 \delta(\tilde{\Omega}_i )\delta(\Gamma_j )
                 \det \mid \{\tilde{\Omega}_i,\Gamma_j \} \mid
                 e^{i S^*},\nonumber \\
        S^*  &=&  \int d^2x \left(
                \pi^\mu {\dot A}_\mu
                + \pi_\phi {\dot \phi}
                + \pi_\theta {\dot \theta}
                + \pi_\rho {\dot \rho} - \tilde{\cal H}^*
                   \right)
\end{eqnarray}
We first discuss about the original unitary gauge
fixing condition as follows
\begin{equation}
\Gamma_i= \Phi^i=(\frac{1}{e}\theta, 
   -\rho, -\pi_\rho, e \pi_\theta ).
\end{equation}
Note that this gauge fixing is consistent 
because when we take the gauge fixing condition 
$\theta \approx 0$, the condition $\rho \approx 0$ is naturally 
generated from the time evolution of $\theta $, i.e., 
$\dot\theta =\{\theta, \tilde{H}^* \}=-e \rho \approx 0 $ ,
and $\pi_\rho$ and $\pi_\theta$ are also successively generated 
from the time evolution of $\rho$ and $\pi_\rho$, respectively.
Choosing the gauge (32), the partition function is found to be
\begin{eqnarray}
Z_U &=&  \int  {\cal D} A_\mu
               {\cal D} \phi
                 e^{i S_U},\nonumber \\
S_U &=&  \int d^2x~ \left[
                \frac{1}{2}\partial_{\mu}\phi\partial^{\mu}\phi
                +eA_{\nu}(\eta^{\mu \nu}
                -\epsilon^{\mu \nu})\partial_{\mu}\phi
        +\frac{1}{2}e^{2}A_{\mu}A^{\mu}~\right],
\end{eqnarray}
which is the same result as in the original phase space expressed by
the Senjanovic path integral formula [19].
Note that although the Maxwell term appearing in (1) is absent in (33)
due to the constraint
$\Omega_3$ in (4), 
the constraints arising from the action (33)
are identical to those obtained from (1) and
this situation coincides with the previous work [7].

Now, let us return to the partiton function (31).
To obtain the partition function in the configuration space,
we perform the momentum integrations by taking the proper order for a 
simpler calculation without any loss of generality.
The $\pi^0$, $\pi_\theta$, and $\pi_\rho$ integrations are trivially
performed by exploiting
the delta functions
$\delta(\widetilde{\Omega}_1)~ =~ \delta[\pi^0 - e\theta]$,
$\delta(\widetilde{\Omega}_4)~ =~ \delta[-e^3 \pi_\phi -e^3 
                \partial_1 \phi +e^4 A_0 - 2e^4 A_1
                - e^4 \rho -e^3 \pi_\theta + e^3 \partial_1 \theta]$,
and $\delta(\widetilde{\Omega}_3)~ =~ \delta[e^2 \pi^1+ e^3 \theta -
e^2 \pi_\rho]$, respectively.
Then, after exponentiating the remaining delta function 
$\delta(\widetilde{\Omega}_2)~ =~ \delta[\partial_1 {\pi}^1
       + e \pi_\phi + e \partial_1 \phi + e^2 A_1 + e^2 \rho ]$ 
with Fourier variable $\xi$ as $\delta(\widetilde{\Omega}_2)~ =~
\int {\cal D} \xi e^{-i\int d^2x~\xi\tilde{\Omega}_2}$, 
we obtain the partition function
\begin{eqnarray}
    Z_{I} &=&  \int  {\cal D} A_\mu  {\cal D} \xi {\cal D} \pi^1
               {\cal D} \phi {\cal D} \pi_\phi
               {\cal D} \theta 
                           {\cal D} \rho   
                 \prod_{j = 1}^{4}
                 \delta(\Gamma_j )
                 \det \mid \{\tilde{\Omega}_i,\Gamma_j \} \mid
                 e^{iS_{I}},\nonumber \\
        S_{I}  &=&  \int d^2x [
                \pi^1 {\dot A}_1
                + \pi_\phi {\dot \phi}
                + {\dot \theta} (- \pi_\phi - \partial_1 \phi - 2e A_1
                + \partial_1 \theta - e \rho ) \nonumber\\
&&     + {\dot \rho} (\pi^1+ e \theta )
                - \frac{1}{2} (\pi_\phi + \frac{1}{e} \partial_1 {\pi}^1 
                + \partial_1 \theta )^2
        - \frac{1}{2} (\partial_1 \phi + \frac{1}{e} \partial_1 {\pi}^1 
                + \partial_1 \theta )^2 \nonumber\\
&&     -( e A_1 -\frac{1}{2} \partial_1 {\pi}^1  
                 -2 \partial_1 \theta + e \rho )
                (\pi_\phi + \partial_1 \phi + e A_1
                + \frac{1}{e} \partial_1 {\pi}^1 + e \rho ) \nonumber\\
&&     + (A_0 -\xi)
                   \widetilde{\Omega}_2                  ] .
\end{eqnarray}
Note that in the usual way $A_0 \rightarrow A_0+\xi$ of the evaluation of the
above partition function, the unwanted $\xi$ field cannot be removed in the measure part of the partition function. 
However, we will show that this problem 
can be resolved by the applying the Fujiwara's technique of the covariant path integral evaluation [3, 15].

The first step is to perform the path integral over $A_0$ 
with the gauge fixing condition
\begin{equation}
\Gamma_1=A_0 \approx 0 .
\end{equation}
The second step is to recover $A_0$ formally by 
identifying the auxiliary field $\xi$ with $-A_0$ field.
Then, the resultant partition function is given by
\begin{eqnarray}
    Z_{II} &=&  \int  {\cal D} A_\mu  {\cal D} \pi^1
               {\cal D} \phi {\cal D} \pi_\phi
               {\cal D} \theta 
                           {\cal D} \rho   
                 \prod_{j = 2}^{4}
                 \delta(\Gamma_j )^{'}
           \det \mid \{\tilde{\Omega}_i,\Gamma_j \} ^{'} \mid
                 e^{iS_{II}},\nonumber \\
        S_{II}  &=&  \int d^2x~ [~
                \pi^1 {\dot A}_1
                + \pi_\phi {\dot \phi}
                + {\dot \theta} (- \pi_\phi - \partial_1 \phi - 2e A_1
                + \partial_1 \theta - e \rho ) \nonumber\\
&&     + {\dot \rho} (\pi^1+ e \theta )
                - \frac{1}{2} (\pi_\phi + \frac{1}{e} \partial_1 {\pi}^1 
                + \partial_1 \theta )^2
        - \frac{1}{2} (\partial_1 \phi + \frac{1}{e} \partial_1 {\pi}^1 
                + \partial_1 \theta )^2 \nonumber\\
&&     -( e A_1 -\frac{1}{e} \partial_1 {\pi}^1  
                 -2 \partial_1 \theta + e \rho )
                (\pi_\phi + \partial_1 \phi + e A_1
                + \frac{1}{e} \partial_1 {\pi}^1 + e \rho ) \nonumber\\
&&     + A_0 \widetilde{\Omega}_2  ~]                 .
\end{eqnarray}
The determinant of $4 \times 4$ matrix in the measure is reduced 
into that of $3 \times 3$ and the prime terms denote that the condition 
$\Gamma_1=A_0 \approx 0$ is imposed after calculations.
Note that the obtained partition function (36) does not contain the $\xi$ field any more.

Next, we perform the Gaussian integration over
$\pi_\phi$ and $\pi^1$ by considering the FP type gauge [7-14]. The 
partition function takes the form
\begin{eqnarray}
    Z_{III} &=&  \int  {\cal D} A_\mu  
               {\cal D} \phi  
               {\cal D} \theta 
               {\cal D} \rho ~  
                 \delta (\dot{A}_1 - \partial_1 A_0
               + \dot{\rho}       )
                 \prod_{j = 2}^{4}
                 \delta(\Gamma_j )^{'}
           \det \mid \{\tilde{\Omega}_i,\Gamma_j \} ^{'} \mid
                 e^{iS_{III}},\nonumber \\
S_{III} &=& \int d^2x  \left[
-\frac{1}{4}F_{\mu \nu}F^{\mu \nu}
              + \frac{1}{2} \partial_\mu \phi \partial^\mu \phi
               + \frac{1}{2} e^2 A_\mu A^\mu
         + e(\eta^{\mu\nu}-\epsilon^{\mu\nu}) 
           A_\nu \partial_\mu \phi \right.
                  \nonumber \\
     && + e \theta (\dot{A}_1 -\partial_1 A_0 ) +e \theta \partial_{\mu} A^{\mu} 
        -e (\partial_1 \phi + e A_1 + \dot{\phi} +\dot{\theta}
        -\partial_1 \theta)\rho \nonumber \\
    &&  \left.-\frac{1}{2} e^2 \rho^2 -\frac{1}{2} \dot{\rho}^2 
        -\frac{1}{2} (\partial_1 \theta)^2 +\frac{1}{2}\dot{\theta}^2
        -\dot{\phi} \dot{\theta} +\partial_1 \phi  \partial_1 \theta        \right].
\end{eqnarray}
The non-trivial $\delta$-function in the measure comes  
from $\int {\cal D} \pi^1 e^{i \int d^2 \! x ~\pi^1 [ F_{01} +\dot{ \rho}] }$.
Due to this $\delta$ function in the measure part,
it is difficult to explicitly show that the above theory has the prevailing gauge symmetry as in the previous works[7,11].
Therefore, to obtain the desired theory,
we must treat this $\delta$ function very carefully.
We'll again make use of the Fujiwara's technique[3] to deal
with this non-trivial $\delta$ function.
First, we choose the second gauge 
\begin{equation}
\Gamma_2=\theta \approx 0,
\end{equation}
and perform the path integral over $\theta$.
After that, we exponentiate the non-trivial 
$\delta$ function with the Fourier variable $\xi '$
as $\int {\cal D} \xi ' e^{-i \int d^2 \! x~ \xi' [ F_{01} + 
{\dot \rho} ]  }$
and recover $\theta$ by 
making the identification $\xi'$ with $e\theta$.
Then, we obtain the following final partition function
\begin{eqnarray}
    Z_{F} &=&  \int  {\cal D} A_\mu  
               {\cal D} \phi  
               {\cal D} \theta 
               {\cal D} \rho   
                 \prod_{j = 3}^{4}
                 \delta(\Gamma_j )^{''}
           \det \mid \{\tilde{\Omega}_i,\Gamma_j \} ^{''} \mid
                 e^{iS_{F}},\nonumber \\
    S_{F}       &=& S_{CSM} + S_{WZ} + S_{NWZ} ~;      \nonumber \\
    S_{CSM} &=&  \int d^2x~ \left[
                -\frac{1}{4}F_{\mu \nu}F^{\mu \nu}
                +\frac{1}{2}\partial_{\mu}\phi\partial^{\mu}\phi
                +eA_{\nu}(\eta^{\mu \nu}
                -\epsilon^{\mu \nu})\partial_{\mu}\phi
        +\frac{1}{2}e^{2}A_{\mu}A^{\mu}~\right], \nonumber \\
    S_{WZ} &=&  - \int d^2x  e \theta ^*F, \nonumber \\
    S_{NWZ} &=&  \int d^2x
            \left[ - e (\partial_1 \phi + eA_1 + \dot\phi -\dot\theta )\rho
                - \frac{e^2}{2} \rho^2 
                - \frac{1}{2}(\dot\rho )^2
                     \right].
\end{eqnarray}
Note that the non-trivial $\delta$ function
from the $\pi^1$
integration in the Liouville measure is finally disappeared.
The double prime terms denotes that $\Gamma_2 =\theta \approx 0$ 
is imposed as well as $\Gamma_1=A_0 \approx 0$ after calculation.
The resultant action has not only the
well-known WZ term $S_{WZ}$ canceling the gauge anomaly and hence being just the $1-cocycle$,
but also a new type of WZ term $S_{NWZ}$,
which is irrelevant to the gauge symmetry but is needed to
make the second class system into the fully first class one
analogous to the case of the CS model [8,10]: In other words,
the final action (39) produces the fully first class constraints system
\begin{eqnarray}
&&\Theta_1 = \pi_0 ,\nonumber \\
&&\Theta_2 =\partial_1 \pi^1 +e \pi_\phi +e \partial_1 \phi +e^2 A_1 +e^2 \rho, \nonumber \\
&&\Theta_3 =-e^2 \theta -e \pi^1 +e \pi_{\rho}, \nonumber \\
&&\Theta_4 =\pi_{\theta}-e \rho, \nonumber \\
&& \{ \Theta_i, \Theta_j \} =0
\end{eqnarray}
(the slight differences from (18) would be the result of the several 
redefinations in the process of evaluation )
with the corresponding Hamiltonian
\begin{eqnarray}
H_F =H_c +\int dx \left[ e \pi_{\phi} \rho -\frac{1}{2} \pi^2_{\rho}
+\frac{e^2}{2} \theta^2 +e \rho (2 e A_1 -e A_0)+ e \theta \pi^1 +e^2 \rho^2
+e \rho \partial _1 \phi \right]
\end{eqnarray}
although the action without $S_{NWZ}$ term does not produce fully first class
constraints but only $det\{\Theta_i, \Theta_j \}=0~(i,j=1~4)$ consistently
with the starting Hamiltonian point of view. Furthermore, the final action
is invariant under the extended gauge transformations as
\begin{eqnarray}
\delta A_\mu = \partial_\mu \Lambda,~\delta \phi = -e \Lambda, 
~\delta \theta = -e \Lambda,~\delta \rho=0, 
\end{eqnarray} 
where the $S_{NWZ}$ in itself is invariant under the above transformation,
which means that this term is not related to the gauge symmetry. 
The appearance of the gauge symmetry in the final action 
is contrast to the previous works [7, 11], where the symmetry is 
not manifest. On the other hand, the lack of the manifest Lorentz 
invariance in $S_F$ does not mean the actual non-invariance of the theory.
This can be proved by the theorem proving the gauge 
independence of the partition function [2, 16].
By choosing the unitary gauge $\rho$ as the third gauge $\Gamma_3$
in the Eq. (37), the Lorentz non-invariant $S_{NWZ}$ term is removed from the 
final action, and thus the Lorentz invariance is recovered.
Consequently the original $gauge~ theory$ is compatible with Lorentz
invariance.
By the way, it is important to note that the final action can not be reduced
to the original starting action by taking the last gauge condition $\Gamma_4$
as $\theta$ because this $\theta$ is really $\xi'/e$ and is different from the 
starting auxiliary field $\theta$ in Eq. (26) which is removed in $\Gamma_3$ and 
$\Gamma_4$ by choosing gauge (37) although this is possible at the beginning step (30) before momentum
$\pi_{\theta}$ and $\pi_{\rho}$ integration.

Finally, we note that our final result (39) does not depend on the particular gauge 
condition $\Gamma_1$ and $\Gamma_2$ due to the gauge independence of the partition 
function (29) [2, 16].

\section{Conclusion}

We have quantized the $a=1$ non-trivial case of the bosonized 
CSM by using the improved BFT formalism, which has provided
the much simpler and transparent insight to the usual BFT method.
We have directly obtained the first class involutive Hamiltonian
from the canonical Hamiltonian by just replacing the original fields with the BFT physical variables,
and reconformed the same relation between the DB defined 
in the original second class system and the Poisson Brackets
defined in the extended phase space as the $a>1$ case.
Furthermore, through the path integral quantization, we have
found the corresponding first class quantum Lagrangian
including a new type of WZ term as well as the well-known WZ term
by resolving the problem of non-trivial delta function, which is only present for the CSM with $a=1$, 
and the unwanted Fourier parameter $\xi$ problem in the measure part. 
Similarly, for the $a>1$ case we could also easily resolve the
$\xi$-problem in our previous result [14] 
by applying the same technique used in this paper.

Through further investigation, 
it will be interesting to apply this newly
improved BFT method to non-Abelian cases as well as an Abelian
four-dimensional anomalous chiral gauge theory, [18]
which seem to be very difficult to analyze within the framework of 
the original BFT formalism [7--11]. Especially for the latter case, we expect that the results, 
including the the appearance of NWZ term, will be very similar to that of the $a=1$ of CSM 
due to the very similar form of the gauge anomaly and constraint structure [18, 20, 21]. 

\section*{Acknowledgments}

We would like to thank 
Y.-W. Kim for helpful discussions. The present study was 
supported by the Basic Science Research Institute Program,
Ministry of Education, 1997, Project No. {\bf BSRI}-97-2414.


\end{document}